\documentclass[aps,prl,twocolumn,showpacs,10pt,nofootinbib,superscriptaddress]
{revtex4-1}

\usepackage{graphicx}
\usepackage[ulem=normalem]{changes}
\usepackage[hang]{subfigure}
\usepackage{nicefrac}
\usepackage[abs]{overpic}
\usepackage{amsmath}
\usepackage{subfigure}
\usepackage{todonotes}
\usepackage{rotating}

\begin{document}

\title{First Measurement of Collectivity of Coexisting Shapes based on Type II Shell Evolution: The Case of $^{96}$Zr}

\author{C. Kremer}
\affiliation{Institut f\"ur Kernphysik, Technische Universit\"at Darmstadt, 
D-64289 Darmstadt, Germany}
\author{S. Aslanidou}
\affiliation{Institut f\"ur Kernphysik, Technische Universit\"at Darmstadt, 
D-64289 Darmstadt, Germany}
\author{S. Bassauer}
\affiliation{Institut f\"ur Kernphysik, Technische Universit\"at Darmstadt, 
D-64289 Darmstadt, Germany}
\author{M. Hilcker}
\affiliation{Institut f\"ur Kernphysik, Technische Universit\"at Darmstadt, 
D-64289 Darmstadt, Germany}
\author{A. Krugmann}
\affiliation{Institut f\"ur Kernphysik, Technische Universit\"at Darmstadt, 
D-64289 Darmstadt, Germany}
\author{P. von Neumann-Cosel}
\affiliation{Institut f\"ur Kernphysik, Technische Universit\"at Darmstadt, 
D-64289 Darmstadt, Germany}
\author{T. Otsuka}
\affiliation{Department of Physics, University of Tokyo, Hongo, Bunkyo-ku, Tokyo 113-0033, Japan}
\affiliation{Center for Nuclear Study, University of Tokyo, Hongo, Bunkyo-ku, Tokyo 113-0033, Japan}
\affiliation{National Superconducting Cyclotron Laboratory, Michigan State University, East Lansing, MI 48824, USA}
\affiliation{Instituut voor Kern- en Stralingsfysica, KU Leuven, B-3001 Leuven, Belgium}
\author{N. Pietralla}
\affiliation{Institut f\"ur Kernphysik, Technische Universit\"at Darmstadt, 
D-64289 Darmstadt, Germany}
\author{V. Yu. Ponomarev}
\affiliation{Institut f\"ur Kernphysik, Technische Universit\"at Darmstadt, 
D-64289 Darmstadt, Germany}
\author{N. Shimizu}
\affiliation{Center for Nuclear Study, University of Tokyo, Hongo, Bunkyo-ku, Tokyo 113-0033, Japan}
\author{M. Singer}
\affiliation{Institut f\"ur Kernphysik, Technische Universit\"at Darmstadt, 
D-64289 Darmstadt, Germany}
\author{G. Steinhilber}
\affiliation{Institut f\"ur Kernphysik, Technische Universit\"at Darmstadt, 
D-64289 Darmstadt, Germany}
\author{T. Togashi}
\affiliation{Center for Nuclear Study, University of Tokyo, Hongo, Bunkyo-ku, Tokyo 113-0033, Japan}
\author{Y. Tsunoda}
\affiliation{Center for Nuclear Study, University of Tokyo, Hongo, Bunkyo-ku, Tokyo 113-0033, Japan}
\author{V. Werner}
\affiliation{Institut f\"ur Kernphysik, Technische Universit\"at Darmstadt, 
D-64289 Darmstadt, Germany}
\author{M. Zweidinger}
\affiliation{Institut f\"ur Kernphysik, Technische Universit\"at Darmstadt, 
D-64289 Darmstadt, Germany}

\date{\today}

\begin{abstract}
\begin{description}
\item[Background] Type II shell evolution has recently been identified as a microscopic cause for nuclear shape coexistence.
\item[Purpose] Establish a low-lying rotational band in $^{96}$Zr.
\item[Methods] High-resolution inelastic electron scattering and a relative analysis of transition strengths are used. 
\item[Results] The $B(\text{E}2;0_1^+\rightarrow 2_2^+)$ value is measured and electromagnetic decay strengths of the $2_2^+$ state are deduced.
\item[Conclusions] Shape coexistence is established for $^{96}$Zr. Type II shell evolution provides a systematic and quantitative mechanism to understand deformation at low excitation energies.
\end{description}
\end{abstract}	
	
	
\maketitle

Understanding structural changes in nuclei, e.g. the development of coexisting structures with different shapes, is a topic of great interest \cite{Heyde}. In this context the role of the monopole (and quadrupole) parts of the proton-neutron (p-n) interaction has previously been recognized \cite{Casten}. It has been shown recently that in particular the monopole part of the tensor interaction plays a crucial role for the explanation of shell evolution with varying proton and neutron numbers (type I) \cite{OtsukatypeI} as well as for configuration-dependent shell evolution (type II) \cite{OtsukatypeII}. While type I shell evolution has been studied extensively, both theoretically and experimentally, cases for type II shell evolution are rare. In particular, data on absolute transition rates for electromagnetic nuclear transitions sensitive to the occurrence of type II shell evolution are still lacking. 

Zirconium isotopes show a quick shape phase transition from spherical ground states for $^{90-98}$Zr to deformed ground states in $^{100}$Zr and heavier isotopes \cite{Federman}. The nucleus $^{96}$Zr has a low-lying excited $0^+$ state, which could be deformed, and is suggested as an example for exhibiting type II shell evolution driven by the tensor force. In fact, shape coexistence has been suggested in the heavier isotope $^{98}$Zr \cite{Wu,Heyde} and has recently been reported for the lighter isotope $^{94}$Zr \cite{Chakraborty}, albeit a considerable mixing of the coexisting structures was deduced from the sizeable interstructure E2 transition strengths. To answer the question whether shape coexistence occurs in $^{96}$Zr knowledge of electromagnetic transition rates is of utmost importance.

In order to guide the later discussion we briefly review the main points of type II shell evolution due to the tensor force, as presented in \cite{OtsukatypeII,OtsukatypeIIb}. The effect of the monopole part of the tensor force depends on the spin-orbit coupling of the respective orbitals. In the following we use the standard notation for $j_> = l+s$ and $j_< = l-s$ quantum numbers with spin $s=\nicefrac12$ and angular momentum $l$. The monopole part of the tensor force is attractive between orbitals with different spin-orbit coupling ($j_>-j'_<$ and $j_<-j'_>$) and repulsive for $j_>-j'_>$ and $j_<-j'_<$ interactions. Thus, a proton excitation from a $j_<$ to a $j_>$ orbital leads to a reduction of spin-orbit splitting for certain neutron orbitals and vice versa (cf. Fig. 1 of Ref. \cite{OtsukatypeIIb}). This leads to an increased likelihood for neutrons to occupy $j'_<$ orbitals, which in turn favors occupation of $j_>$ orbitals for protons. This is a self-reinforcing effect, which can stabilize low-lying deformed configurations.

The nucleus $^{96}$Zr is a well-suited candidate for featuring type II shell evolution because lifting protons from $p_{\nicefrac12}$, $f_{\nicefrac52}$ orbitals to the $g_{\nicefrac92}$ orbital affects the occupancy of neutron orbitals.

It is the purpose of this Letter to report on an electron scattering experiment off $^{96}$Zr which determines the transition strengths of the $2_2^+$ state to low-lying states including the first excited $0_2^+$ state. The interpretation of the $0_2^+$ state and the band built on top of it as a deformed structure is confirmed, the deformation is deduced, and type II shell evolution is identified as the main stabilizing mechanism of the deformed excited states.

\begin{figure}
	\vspace{.3cm}
	\centering
	\includegraphics{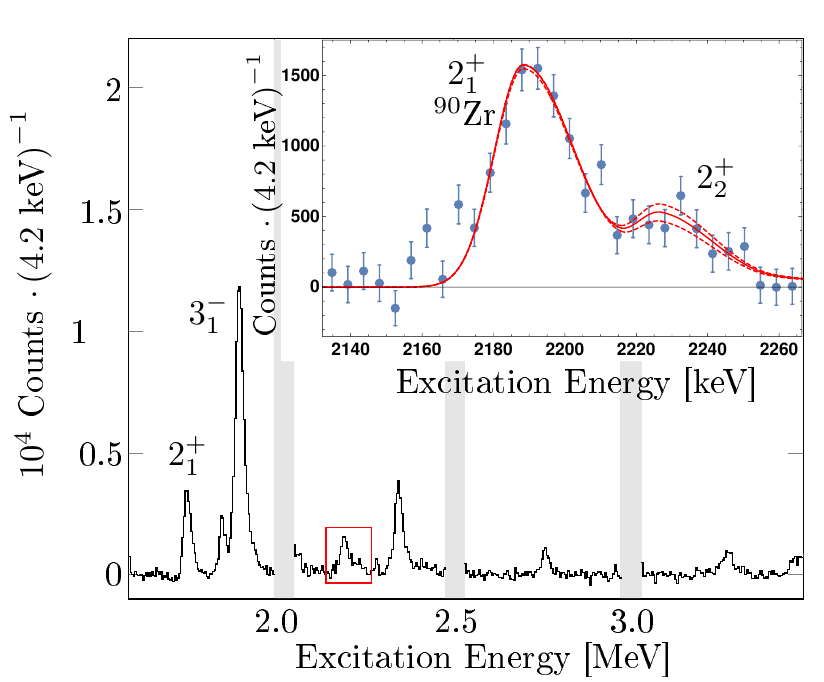}
	\caption{(Color online) Summed and efficiency corrected experimental data for $\theta = 141^{\circ}$ and $E_0~=~43\text{ MeV}$ after the radiative tail of the elastic line has been subtracted. Grey areas correspond to inactive segments of the detector system. The inlet shows a magnification of the region around the $2_2^+$ state of $^{96}$Zr, which is indicated by the red rectangle.}
	\label{fig:sum141deg}
	\end{figure}

The experiment was conducted at the Superconducting DArmstadt LINear ACcelerator (S-DALINAC) using the Lintott high-resolution magnetic spectrometer \cite{Graef}. Data were taken at scattering angles of $81^{\circ}$, $93^{\circ}$, $117^{\circ}$, and $141^{\circ}$. An electron energy of $43$ MeV was used except for the measurement at $117^{\circ}$, which was performed at $E_0 = 69$ MeV. The covered momentum transfer ($q$) values were $q = 0.59\,\text{fm}^{-1}$, $0.40\,\text{fm}^{-1}$, $0.31\,\text{fm}^{-1}$, and $0.28\,\text{fm}^{-1}$. Intensities of the beam ranged from $0.5~\mu\text{A}$ up to $2.5~\mu\text{A}$ and were limited by the dead time of the data acquisition system \cite{Lenhardt}.
The target used was a $2$x$3$~cm$^2$ self-supporting zirconium foil of thickness $10$ mg$/$cm$^2$. It was enriched in $^{96}$Zr up to $57.36\,\%$ and also contained $^{92}$Zr ($27.2\,\%$), $^{90}$Zr ($9.2\,\%$), $^{94}$Zr ($4.3\,\%$), and $^{91}$Zr ($2.0\,\%$). The resolution of the obtained spectra ranged from $12.3$ keV to $17.5$ keV full width at half maximum (FWHM). As an example the spectrum obtained at $141^\circ$ is shown in Fig. \ref{fig:sum141deg}. The $0_1^+ \rightarrow 2_2^+$ transition of $^{96}$Zr is located close to the stronger $0_1^+\rightarrow 2^+_1$ transition of $^{90}$Zr with an energy difference of $22\text{ keV}$ only making good resolution critical to the analysis of this experiment as highlighted by the inlet in Fig. \ref{fig:sum141deg}. 

\begin{figure}
\includegraphics{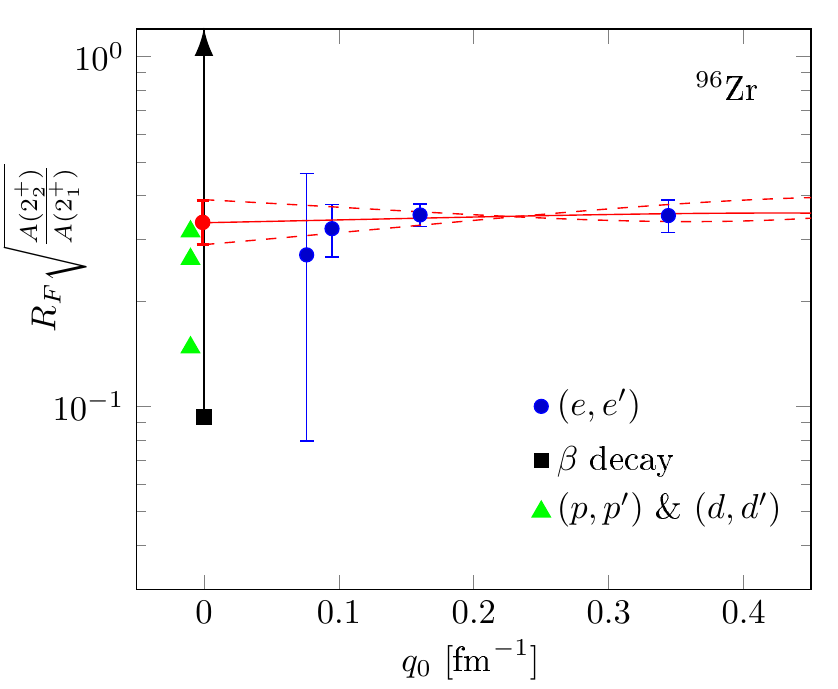}
\caption[Extraction of $\sqrt{\nicefrac{B(\text{E}2;2_2^+\rightarrow 0_1^+)}{B(\text{E}2;2_1^+\rightarrow 0_1^+)}}$ from experimental data.]{(Color online) Value of $R_F\,\sqrt{\nicefrac{A\left(2_2^+\right)}{A\left(2_1^+\right)}}$ as a function of elastic momentum transfer $q_0$. The solid red line shows the best fit of Eq. (\ref{eq:BE2Fit}) to the experimental data (blue circles). The dashed lines represent the $1\sigma$ uncertainties with respect to $T(2_2^+,2_1^+)$. The adopted literature lower limit from $\beta$ decay \cite{Mach} is shown as a black square. The green triangles, shifted slightly to the left for readability, represent estimates from $(p,p')$, $(d,d')$, and polarized $(d,d')$ measurements \cite{Hofer}.}
\label{fig:Result}
\end{figure}

The experimental raw data, consisting of many single runs, were efficiency corrected, energy calibrated, and summed. Then, the elastic background was removed assuming identical lineshapes for the peaks corresponding to elastic and inelastic scattering. The number of detected counts $A_i$ for each excited state $i$ can be determined by a $\chi^2$ minimization using an empirical line shape tailored to electron scattering experiments \cite{Hofmann}. The extracted peak areas allow to determine the strength of the $0_1^+\rightarrow 2_2^+$ transition relative to that of the $0_1^+\rightarrow 2_1^+$ transition using a Plane Wave Born Approximation (PWBA) \cite{Scheikh1,Scheikh2}  (cf. Fig. \ref{fig:Result}). The Coulomb corrections accounting for the distortion of the electron wave functions by the nucleus cancel in this relative analysis to better than $1\,\%$ over the momentum transfer range of interest. Employing the PWBA formalism using Siegert's theorem and expanding the transition strengths $B(\text{C}\,\lambda,q)$ in powers of the momentum transfer ($q$) and transition radius $(R_{tr})$ yields a relation of experimentally determined peak areas to the ratio of $B(\text{E}2)$ values
\begin{widetext}
\begin{equation}
			R_F(q)\,\sqrt{\frac{A_2}{A_1}}\approx\sqrt{\frac{B(E2, k_2)}{B(E2, k_1)}}\cdot\left(\frac{1-\frac{q_2^2}{14}\,(R_{tr,1}+\Delta R)^2+\frac{q_2^4}{504}\,(R_{tr,1}+\Delta R)^4}{1-\frac{q_1^2}{14}\,(R_{tr,1})^2+\frac{q_1^4}{504}\,(R_{tr,1})^4}\right),
\label{eq:BE2Fit}
\end{equation}
\end{widetext}
where $R_F(q)$ denotes a ratio of kinematical factors \cite{Scheikh1,Scheikh2}, $R_{tr,1}$ is the transition radius of the $2_1^+$ state, $\Delta R = R_{tr,2} - R_{tr,1}$, and $k_1$ and $k_2$ denote the photon point momentum transfers for the excitation of $2_1^+$ and $2_2^+$ states. A $\chi^2$ minimization of Eq. (\ref{eq:BE2Fit}) with respect to $T(2_2^+,2_1^+) = \sqrt{B(\text{E}2;2_2^+\rightarrow 0_1^+)/B(\text{E}2;2_1^+\rightarrow 0_1^+)}$ and the difference in transition radii $\Delta R$ is carried out. The best fit is represented by the solid red line in Fig. \ref{fig:Result} whereas the dashed lines represent the solutions that define the $1\,\sigma$ uncertainties for $T(2_2^+,\,2_1^+)$. 
Using a transition radius $R_{tr,1}=5.38\text{ fm}$, taken from a QRPA calculation, this analysis yields
\begin{eqnarray*}
	T(2_2^+,2_1^+) &=& 0.34 \substack{+0.05 \\ -0.04}\text{, and} \label{eq:ratioResult} \\
	\Delta R &=& \left(-0.22 \substack{+0.87 \\ -0.92}\right)\,\text{fm.}
\end{eqnarray*}
The extracted value of $\Delta R$ is consistent with zero with an uncertainty of about $\pm 1$ fm. However, the extracted value of $T(2_2^+,\,2_1^+)$ is largely independent of the choice of $R_{tr,1}$ (at least up to $\pm 1\text{ fm}$), see e.g. Fig. 5 of Ref. \cite{Scheikh1}. Thus, the data are insensitive to a possible difference of transition radii $\Delta R$ of the $0^+_1 \rightarrow 2^+_{1,2}$ transitions (again to about $\pm 1$ fm) indicating that the determination of $T(2_2^+,2_1^+)$ is independent of $\Delta R$. Combining Eq. (\ref{eq:BE2Fit}) with the literature value $B(\text{E}2;\,2_1^+\rightarrow 0_1^+) = 2.3 \pm 0.3 \text{Weisskopf units (W.u.)}$~\cite{NNDC} yields $B(\text{E}2;\,2_2^+\rightarrow 0_1^+) = 7.4(2.3)~e^2\text{fm}^4 = 0.26(8) \text{ W.u.}$. Together with the multipole mixing ratios and the branching ratios taken from Ref. \cite{NNDC} it is then possible to determine all the transition strengths for electromagnetic decays of the $2_2^+$ state (see Tab. \ref{tab:shellmodel}). For the first time our data provide model-independent finite values from an electromagnetic probe for the decay rates of the $2_2^+$ state of $^{96}$Zr including the E2 decay to the $0_2^+$ state crucial to determine its structure.
The collective value, $B(\text{E}2;\,2_2^+\rightarrow 0_2^+)=36(11)\text{ W.u.}$, hints at a common deformed structure of the $0_2^+$ and $2_2^+$ states. Assuming a rigid, axial symmetric, deformed shape the quadrupole deformation parameter $\beta_2$ can be estimated
\begin{equation}
\beta_{2} = \frac{4\,\pi}{3\,Z\,R_0^2}\cdot\left(\frac{B(\text{E}2;\,0_2^+\rightarrow 2_2^+)}{e^2}\right)^{\frac12} \approx 0.24,
\end{equation}
where $R_0 = 1.2 \cdot A^{\nicefrac13}\text{ fm}$ has been used. Thus the collective $B(\text{E}2;\,2_2^+\rightarrow 0_2^+)$ value indicates well deformed $0^+_2$ and $2_2^+$ states while the weak $B(\text{E}2;\,2_1^+\rightarrow 0_1^+)$ strength indicates a nearly spherical ground state for $^{96}$Zr.


In light of the experimental data obtained in this work a new shell model calculation for $^{96}$Zr has been performed. The model space consists of $1f_{\nicefrac52}$, $2p_{\nicefrac32}$, $2p_{\nicefrac12}$ orbitals, and the full $sdg$ shell for protons and the full $sdg$ shell, plus $1h_{\nicefrac{11}{2}}$, $3p_{\nicefrac32}$, and $2f_{\nicefrac72}$ orbitals for neutrons. This model space is considerably larger than that of previous shell model calculations (see, e.g., \cite{Sieja}). Details of the shell model calculation can be found in the preceding Letter \cite{Togashi}. A comparison of transition strengths between low-lying states to the experimental values using effective charges $e_p = 1.3\,e$ and $e_n = 0.6\,e$ is shown in Tab. \ref{tab:shellmodel}.

\begin{table}[htb]
\centering
\caption[Comparison transition strengths for the low-lying states of $^{96}$Zr with the shell-model calculations.]{Comparison of transition strengths for the low-lying states of $^{96}$Zr to the shell-model calculations (SM) and a two-state model with (TSM$_m$) and without mixing (TSM$_u$) described in the text. Experimental data obtained in this work are marked by an asterisk (*).}
\begin{tabular}{lcccc} \hline
& experiment & SM & TSM$_u$ & TSM$_m$ \\[.1mm] \hline
$B(\text{E}2;2_1^+\rightarrow 0_1^+)$ [W.u.] & $2.3(3)$ &  1.28 & 2.5 & 2.3\\[.5mm]
$B(\text{E}2;2_2^+\rightarrow 0_2^+)$ [W.u.] & $36(11)^*$ & 52.7 & 36.7 & 36\\[.5mm]
$B(\text{E}2;2_2^+\rightarrow 0_1^+)$ [W.u.] & $0.26(8)^*$ & 0.00 & 0.00 & 0.26\\[.5mm]
$B(\text{M}1;2_2^+\rightarrow 2_1^+)$ [$\mu_N^2$] & $0.14(5)^*$ & 0.01 & 0.00 & 0.07\\[.5mm]
$B(\text{E}3;3_1^-\rightarrow 0_1^+)$ [W.u.] & 57(4) & 37.3 & - & -\\[.5mm] 
$B(\text{E}1;2_2^+\rightarrow 3_1^-)$ [W.u.] & $28(9)\cdot 10^{-3~*}$ & 0.00 & - & -\\[.5mm]\hline
\end{tabular}
\label{tab:shellmodel}
\end{table} 

The enhanced $2_2^+\rightarrow 0_2^+$ transition and the small $2_1^+\rightarrow 0_1^+$ transition strength calculated within the shell model are in qualitative agreement with the experiment. The strong octupole collectivity, a hallmark of $^{96}$Zr, is reproduced within $20~\%$. The shell model does not describe the finite $B(M1)$ value between the $2^+$ states, which originates from a delicate mixing of the two configurations and hints at a small but finite mixing of the spherical and deformed states. 

\begin{figure}
	\includegraphics{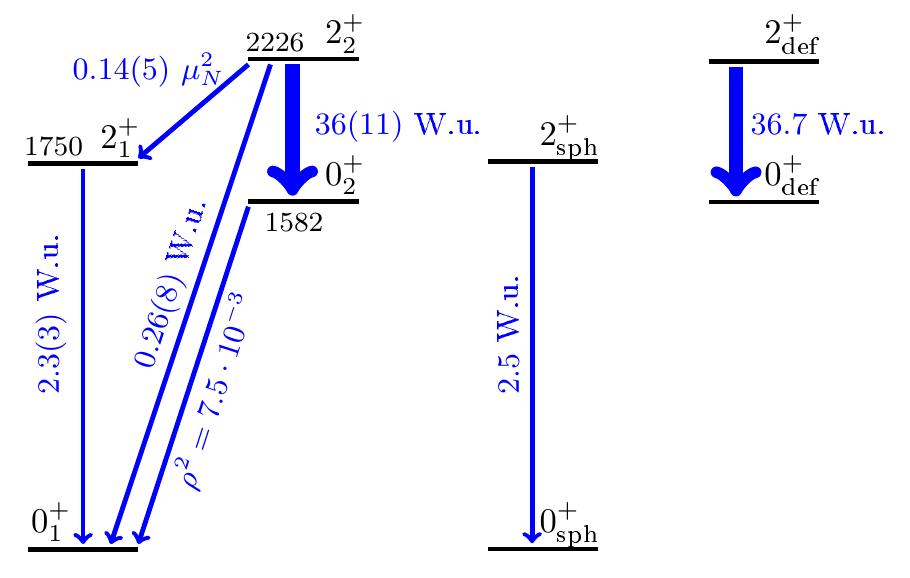}
\caption[Low-lying states of $^{96}$Zr (left) and assumed underlying structure (right).]{(Color online) Low-lying $0^+$ and $2^+$ states of $^{96}$Zr (left) and assumed underlying structure (right). Energies are given in keV. Note that only the $B(\text{E}2;2_1^+\rightarrow 0_1^+)$ and $B(\text{E}2;2_2^+\rightarrow 0_2^+)$ transition strengths have been used in the mixing calculation.}
\label{fig:MixingSchematics}
\end{figure}

As both, a deformed and a spherical configuration, coexist at low energies, it is instructive to look at the data in a two-state model (TSM) analysis. Assuming that the experimentally observed states (l.h.s. of Fig. \ref{fig:MixingSchematics}) are mixtures of deformed and spherical structures their wave functions can be written as
\begin{equation}
\begin{array}{r@{}l}
\mid 0^+_1 \rangle &{}= \alpha\mid 0^+_{\text{sph}} \rangle + \beta\mid 0^+_{\text{def}} \rangle \\
\mid 0^+_2 \rangle &{}= -\beta\mid 0^+_{\text{sph}} \rangle + \alpha\mid 0^+_{\text{def}} \rangle \\
\mid 2^+_1 \rangle &{}= \gamma\mid 2^+_{\text{sph}} \rangle + \delta\mid 2^+_{\text{def}} \rangle \\
\mid 2^+_2 \rangle &{}= -\delta\mid 2^+_{\text{sph}} \rangle + \gamma\mid 2^+_{\text{def}} \rangle
\end{array}
\end{equation}
where $\alpha$, $\beta$, $\gamma$, and $\delta$ are amplitudes normalized to $\alpha^2+\beta^2 = \gamma^2+\delta^2 = 1$. Using the experimental excitation energies of these states and the observed transition strengths $B(\text{E}2;\,2^+_1\rightarrow 0^+_1)$ and $B(\text{E}2;\,2^+_2\rightarrow 0^+_2)$ as input data, the mixing amplitudes can be computed under the assumption that the mixing matrix element $V_{mix}$ between the spherical and deformed structures is identical for the $0^+$ and $2^+$ states and E$2$ matrix elements between pure configurations vanish. Carrying out the calculation yields
\begin{eqnarray*}
\alpha^2 &= 99.8\,\% \quad \beta^2 &= 0.2\,\% \\
\gamma^2 &= 97.5\,\% \quad \delta^2 &= 2.5\,\%.
\end{eqnarray*}
Thus, the states are decoupled to a very good approximation, which is also supported by $\beta$ decay data \cite{Mach2}. The mixing matrix element amounts to $V_{mix} = 76\text{ keV}$. The interband $B(\text{E}2)$ values of the unmixed configurations (r.h.s. of Fig. \ref{fig:MixingSchematics}) are almost identical to the mixed case. 

By measuring the electromagnetic decay properties we have established the high purity of the coexisting states. The present shell model interaction catches the dominant components of the wave functions but is incapable in describing their weak mixing on a percent level. The TSM also provides information on them and hence can be compared to the shell model results. In this case it is justified to interpret the shell model states as an approximation to the pure states of the TSM (cf. r.h.s of Fig. \ref{fig:MixingSchematics}. Mixing the shell model states with the deduced mixing matrix element $V_{mix}$ leads to $B(\text{M1};2_2^+\rightarrow 2_1^+) \approx 0.07 \mu_N^2$, where the rotational model $g$ factor $(\approx Z/A)$ has been used to determine the matrix element $\langle 2^+_{def}\mid \text{M1} \mid 2^+_{def}\rangle$ and the corresponding matrix element of the $2^+_{sph}$ state has been calculated using the $g$ factor of the $2^+_1$ state ($-0.256\,\mu_N^2$) given by the shell model calculation. The resulting $B(\text{M1};2_2^+\rightarrow 2_1^+)$ value of $0.07~\mu_N^2$ is reasonably close to the experimental value (Tab. \ref{tab:shellmodel}), which again suggests that the different shapes coexist with little mixing. Type II shell evolution has been established as the underlying mechanism for stabilizing such a structure.

To gain further insight into this mechanism we turn to the shell-model results in terms of occupation numbers summarized in Tab. \ref{tab:occupationNumbers}.
\begin{table}
\centering
\begin{tabular}{lcccccccc} \hline
 & \multicolumn{2}{c}{$0_1^+$} & \multicolumn{2}{c}{$2_1^+$} & \multicolumn{2}{c}{$0_2^+$} & \multicolumn{2}{c}{$2_2^+$} \\
& $\pi$ & $\nu$ & $\pi$ & $\nu$ & $\pi$ & $\nu$ & $\pi$ & $\nu$ \\ \hline
$2f_{\nicefrac72}$ & - & 0.09 & - & 0.06 & - & 0.12 & - & 0.08 \\
$3p_{\nicefrac32}$ & - & 0.01 & - & 0.01 & - & 0.01 & - & 0.01 \\
$1h_{\nicefrac{11}{2}}$ & - & 0.68 & - & 0.40 & - & 2.04 & - & 1.96\\
$1g_{\nicefrac72}$ & 0.14 & 0.09 & 0.11 & 0.06 & 0.15 & 1.38 & 0.15 & 1.37 \\
$2d_{\nicefrac32}$ & 0.06 & 0.10 & 0.04 & 0.16 & 0.05 & 0.71 & 0.05 & 0.73 \\
$3s_{\nicefrac12}$ & 0.01 & 0.20 & 0.01 & 1.11 & 0.03 & 0.41 & 0.03 & 0.38 \\
$2d_{\nicefrac52}$ & 0.08 & 5.04 & 0.07 & 4.38 & 0.22 & 1.69 & 0.21 & 1.84 \\
$1g_{\nicefrac92}$ & 0.51 & 9.79 & 0.42 & 9.81 & 3.49 & 9.64 & 3.40 & 9.63 \\
$2p_{\nicefrac12}$ & 1.85 & - & 1.87 & - & 0.80 & - & 0.88 & - \\
$2p_{\nicefrac32}$ & 3.59 & - & 3.69 & - & 2.54 & - & 2.60 & - \\
$1f_{\nicefrac52}$ & 5.76 & - & 5.79 & - & 4.72 & - & 4.68 & - \\ \hline
\end{tabular}
\caption{Occupation numbers for the full model space of the shell model calculation. A dash marks an orbital that is not part of the respective model space.}
\label{tab:occupationNumbers}
\end{table}
For the ground state of $^{96}$Zr all orbitals below the $Z = 40$ and $N = 56$ subshell closures are, to good approximation, filled and those above the subshell closures are empty. The structure of the $2_1^+$ state is similar to that apart from one neutron excited from the $2d_{\nicefrac52}$ orbital to the $3s_{\nicefrac12}$ orbital. The deformed $0_2^+$ and $2_2^+$ states are also very similar to one another, but markedly different from the spherical states. On average three protons are excited from the $pf$ shell to the $1g_{\nicefrac92}$ orbital. In addition a total of three neutrons are excited from the $2d_{\nicefrac52}$ and the $3s_{\nicefrac12}$ orbitals to the $2d_{\nicefrac32}$, $1g_{\nicefrac72}$, and $1h_{\nicefrac{11}{2}}$ orbitals. The large fragmentation of the resulting wave function in terms of spherical shell-model components is indicative of deformation.

\begin{figure}[htb]
\includegraphics{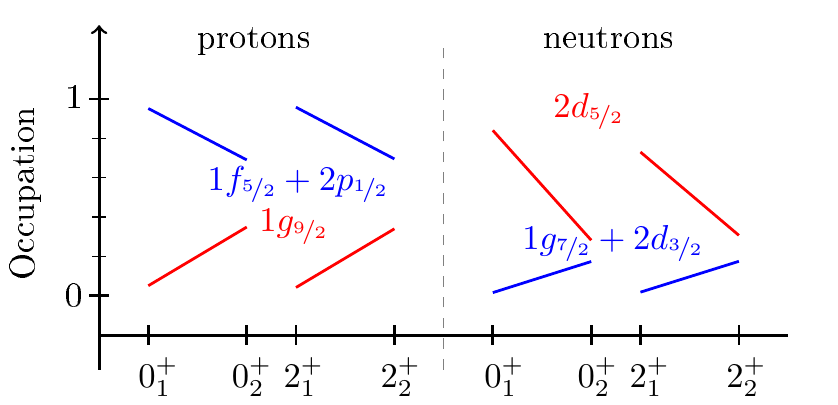}
\caption{(Color online) Illustration of relative occupation of $j_>$ (red) and $j_<$ orbitals (blue) for protons and neutrons. For the deformed states ($0_2^+$, $2_2^+$) protons are more likely to occupy $j_>$ orbitals and neutrons are more likely to occupy $j_<$ orbitals than for the spherical states. This pattern is typical for type II shell evolution. See text for further details.}
\label{Fig:Type2}
\end{figure}

The difference of occupation numbers between spherical and deformed states (Fig. \ref{Fig:Type2}) can be understood in terms of type II shell evolution. For the deformed $0_2^+$ and $2_2^+$ states the protons in the $1g_{\nicefrac92}$ orbital ($j_>$) lead to a reduction of spin-orbit splitting in the neutron sector caused by the monopole part of the tensor force. This effect is enhanced by the fact that the protons are excited to the $1g_{\nicefrac92}$ orbital predominantly from $j_<$ orbitals in the $pf$ shell (l.h.s. of Fig. \ref{Fig:Type2}). The difference in single particle energies between the $2d_{\nicefrac52}$ and $1g_{\nicefrac72}$ orbitals is reduced from $4.0 \text{ MeV}$ for the spherical states ($N = 56$ subshell closure) to $2.1 \text{ MeV}$ for the deformed states \cite{Togashi}. Additionally, the neutron single particle energies of the deformed states are more densely packed around the Fermi energy, which explains the fragmentation of the wave function. The increased occupation number of the $\nu\left(1h_{\nicefrac{11}{2}}\right)$ orbital seems to contradict the above said, however, it is due to the central force which outweighs the effect of the tensor force for the interaction of the $\nu\left(1h_{\nicefrac{11}{2}}\right)$-$\pi\left(1g_{\nicefrac{9}{2}}\right)$ orbitals. In the deformed states the neutrons are more likely to occupy $j_<$ levels than in the spherical states (r.h.s. of Fig. \ref{Fig:Type2}). This in turn leads to an increase in spin-orbit splitting in the proton sector. The present shell model calculation shows that the $2p_{\nicefrac12}$-$1g_{\nicefrac92}$ gap in the proton effective single particle energies is lowered from $3.3 \text{ MeV}$ in the spherical states ($Z = 40$ subshell closure) to approximately $1.2 \text{ MeV}$ in the deformed states. Thus, the self-reinforcing effect of type II shell evolution is evident for the nucleus $^{96}$Zr.

To summarize, electron scattering has been used to measure the $0_1^+\rightarrow 2_2^+$ transition in $^{96}$Zr and determine its strength in a relative PWBA analysis. Using known branching ratios and multipole mixing ratios the electromagnetic decay strengths of the $2_2^+$ state have been deduced. The $2_2^+\rightarrow 0_2^+$ transition strength establishes the $2_2^+$ state as a collective excitation on top of a deformed $0_2^+$ state with deformation parameter $\beta_2\approx 0.24$. Type II shell evolution is identified as the stabilizing mechanism for the shape coexistence of this low-lying deformed state and the band build on top of it. Thus, the nucleus $^{96}$Zr represents the first example of type II shell evolution supported by the measurement of electromagnetic observables.

This work was supported by the Deutsche Forschungsgemeinschaft under Grant No. SFB 1245 and by Grants-in-Aid for
Scientific Research (23244049). It was supported in part by the HPCI Strategic Program (hp150224), in part by the MEXT and JICFuS as a priority issue (Elucidation of the fundamental laws and evolution of the universe) to be tackled by using Post 'K' Computer (hp160211), and by the CNS-RIKEN joint project for large-scale nuclear structure calculations.



\end{document}